\documentclass[sigconf]{acmart}

\usepackage{microtype}
\usepackage{graphicx}
\usepackage{subfigure}
\usepackage{booktabs}
\usepackage{hyperref}
\usepackage{xspace}
\usepackage{enumitem}
\usepackage{caption}
\usepackage[linesnumbered,ruled,vlined]{algorithm2e}

\AtBeginDocument{%
  }

\copyrightyear{2025}
\acmYear{2025}
\setcopyright{cc}
\setcctype{by}
\acmConference[SC '25]{The International Conference for High Performance Computing, Networking, Storage and Analysis}{November 16--21, 2025}{St Louis, MO, USA}
\acmBooktitle{The International Conference for High Performance Computing, Networking, Storage and Analysis (SC '25), November 16--21, 2025, St Louis, MO, USA}
\acmDOI{10.1145/3712285.3759864}
\acmISBN{979-8-4007-1466-5/2025/11}

\newcommand{\proj}{\emph{MLP-Offload}\xspace}

\begin{document}

\author{Avinash Maurya}
\affiliation{
    \institution{Argonne National Laboratory}
    \city{Lemont, IL}
    \country{USA}
}
\email{amaurya@anl.gov}

\author{M. Mustafa Rafique}
\affiliation{
    \institution{Rochester Institute of Technology}
    \city{Rochester, NY}
    \country{USA}
}
\email{mrafique@cs.rit.edu}

\author{Franck Cappello}
\affiliation{
    \institution{Argonne National Laboratory}
    \city{Lemont, IL}
    \country{USA}
}
\email{cappello@mcs.anl.gov}

\author{Bogdan Nicolae}
\affiliation{
    \institution{Argonne National Laboratory}
    \city{Lemont, IL}
    \country{USA}
}
\email{bogdan.nicolae@acm.org}
\renewcommand{\shortauthors}{Avinash Maurya et al.}

\title{\proj: Multi-Level, Multi-Path Offloading for LLM Pre-training to Break the GPU Memory Wall}

\begin{abstract}
Training LLMs larger than the aggregated memory of multiple GPUs is increasingly necessary due to the faster growth of LLM sizes compared
to GPU memory. To this end, multi-tier host memory or disk offloading techniques are proposed by state of art. Despite advanced asynchronous multi-tier read/write strategies, such offloading strategies result in significant I/O overheads in the critical path of training, resulting in slower iterations.  To this end, we propose \proj, a novel multi-level, multi-path offloading engine specifically designed for optimizing LLM training on resource-constrained setups by mitigating I/O bottlenecks. We make several key observations that drive the design of \proj, such as I/O overheads during the update dominate the iteration time; I/O bandwidth of the third-level remote storage tier remains unutilized; and, contention due to concurrent offloading amplifies I/O bottlenecks. Driven by these insights, we design and implement \proj to offload the optimizer states across multiple tiers in a cache-efficient and concurrency-controlled fashion to mitigate I/O bottlenecks during the backward and update phases. Evaluations on models up to 280B parameters shows that \proj achieves 2.5$\times$ faster iterations compared to the state-of-the-art LLM training runtimes.
\end{abstract}



\maketitle

\section{Introduction}
\label{sec:intro}
Large Language Models (LLMs) have revolutionized a broad range of domains thanks to text summarizing and knowledge distillation, enabling researchers to navigate complex scientific literature more efficiently. A natural evolution
towards more general foundational models (FMs) capable of capturing
complex correlations between different data modalities (e.g. using cross-attention) are beginning to unlock an even bigger impact on scientific progress.

In a quest for advancing emergent behavior (capabilities not explicitly trained for but emerging spontaneously due to the massive scale and exposure to vast amounts of data during training), FMs' scale and complexity continuously increase, requiring larger training infrastructures and
incurring enormous costs. Pre-training and even fine-tuning were not even feasible without large-scale HPC systems: ChatGPT-3 was trained on over 10,000 GPUs~\cite{chatGPT-cost} and cost over \$4 million in each training session. Meta’s LLaMA model (released Feb. 2023), follows the same pattern: it used 2,048 Nvidia A100 GPUs to train on 1.4 trillion tokens, taking about 21 days~\cite{CNBC}. Even worse, LLMs are rapidly growing both in terms of parameters and training data sizes. For example, ChatGPT-4 is estimated to be $10\times$ larger and was trained on $570\times$ more data than ChatGPT-3. Under these circumstances, the scale and cost required to train or fine-tune FMs become prohibitively expensive.

\begin{figure}[t]
    \centering
    \includegraphics[width=0.8\linewidth]{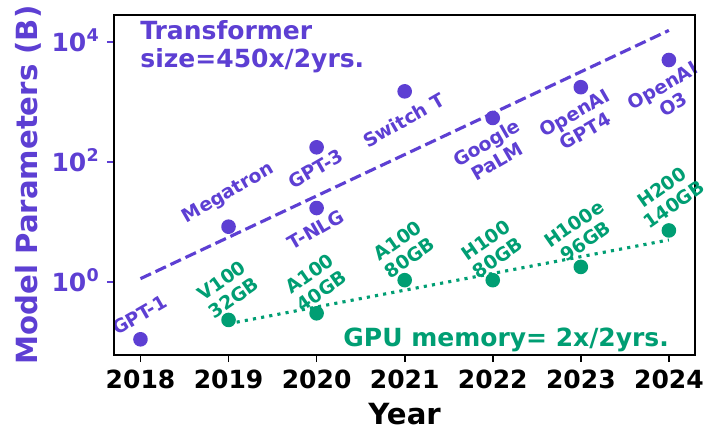}
    \vspace{-10pt}
    \caption{Model vs GPU memory growth.}
    \label{fig:model-vs-hardware}
    \Description{A figure illustrating the growth trend of large language models vs. the GPU memory, wherein the GPU memory grows at a much slower pace, necessitating the need for advanced memory optimizations and parallelization strategies to accomodate and run large models in democratized fashion.}
\end{figure}

\paragraph*{\bf Motivation: GPU Memory Wall}
A large number of state-of-the-art techniques have been proposed to make efficient use of the computational capabilities and massive parallelism
offered by a large number of GPUs at scale. For example, DeepSeek~\cite{liu2024deepseek} has introduced several techniques in this direction. However, with GPU computations becoming more efficient, the bottleneck is shifting towards insufficient GPU memory capacities. For instance, the Gopher-280B model~\cite{rae2022scalinglanguagemodelsmethods} requires 4.8~TB~\cite{ZeroInfinity-SC21}, demonstrating terabyte-scale GPU memory requirements to train models in the order of hundreds of billions of parameters. This bottleneck leads to a so-called \emph{memory wall}: the ever-increasing size of model parameters and auxiliary data structures such as activations and optimizer state (450$\times$/2~years) far exceed the GPU memory growth of newer generations of devices (2$\times$/2~years)~\cite{gholami2024ai,model-sizes,maurya2024breaking}. This trend, illustrated in Figure~\ref{fig:model-vs-hardware} is unsustainable. Furthermore,
a majority of HPC datacenters need to serve multiple users running
different workloads at the same time~\cite{generic-ml-workload-study}. Thus, the ability to train and
fine-tune LLMs and FMs on a limited number of GPUs is highly desirable
compared with the alternative of waiting for a long time in the batch queue
to gain access to a large number of GPUs~\cite{duan2024efficienttraininglargelanguage, jiang2024neosavinggpumemory}.

To address this challenge, state-of-the-art LLM training frameworks (e.g. DeepSpeed~\cite{rasley2020deepspeed} and Megatron~\cite{shoeybi2019megatron}) introduce several techniques such as ZeRO redundancy optimization (e.g. partitioning the large-optimizer state across data-parallel replicas)~\cite{zero-deepspeed-intro}, quantization, mixed-precision training~\cite{shoeybi2019megatron}, etc. Still, this is not enough
to alleviate the large GPU memory requirements. As a consequence, these
techniques are often complemented with \emph{offloading} of important
data structures (model parameters, optimizer states) to the slower host memory (DRAM)~\cite{maurya2024deep}. When memory constraints are even tighter, the aggregated GPU and host memory capacity are not enough, prompting the need to further offload the data structures to third-level storage tiers (e.g., NVMe devices)~\cite{ZeroInfinity-SC21,liao2024adding,ren2021zero}.
While the slower memory and storage tiers have significantly higher capacities, they also have significantly lower I/O bandwidth.
For example, when pre-training a 20B LLaMA model without offloading (model parameters and optimizer states reside on aggregated GPU memory of a single node), the average iteration duration is 0.4~seconds. Offloading the optimizer states, used during the update phase, to the host memory increases the duration of the iterations to 3.7~seconds on average. Further offloading to a node-local NVMe increases the duration of the iterations to 67~seconds~(\S~\ref{sec:motivation}). Such a
large I/O bottleneck results in an overall 170$\times$ slowdown compared with training without offloading. At larger scales, such as training the 70B-parameter LLaMA across multiple nodes, communication overheads begin to dominate iteration time, which in turn amortizes the relative cost of NVMe offloading. As a result, the slowdown of NVMe-offloaded training relative to GPU-only training is reduced by about 7$\times$, highlighting the interaction between communication and I/O bottlenecks in distributed settings.
Therefore, despite the scale, reducing the I/O bottlenecks when using third-level storage is an important challenge that needs to be solved in order to be able to benefit from a significantly higher overall memory capacity without a prohibitive performance penalty.

\paragraph*{\bf Limitations of State-of-the-art}
Existing advanced LLM training runtimes such as Microsoft's DeepSpeed~\cite{rasley2020deepspeed}, PyTorch's Fully Sharded Data Parallel~(FSDP)~\cite{Zhao:FSDP:VLDB2023}, Colossal-AI~\cite{li2023colossal}, etc. introduce novel asynchronous offloading techniques for both model and optimizer states. While FSDP allows only offloading to host memory, DeepSpeed and Colossal-AI runtimes enable NVMe offloading. Specifically, the model parameters and optimizer state are partitioned into model shards, which
are distributed among worker processes (one per GPU) and collaborate
using model parallelism. In turn, each model shard is decomposed into smaller chunks, called ``subgroups'', which are processed in combination with mixed-precision techniques one by one as follows. During the forward pass and backward pass, a copy of the FP16 parameters is used on the GPUs to compute the gradients of each subgroup. As the backward pass progresses, the
gradients are flushed to host memory and from there to third-level
storage in FP32 format. Then, the update phase is performed on the CPU
to avoid excessive traffic between the host memory and GPU memory, which
negates the computational speed-up of GPUs. During the update phase
the FP32 model parameters, optimizer state and gradients of each subgroup
are fetched from third-level storage to host memory, the CPU-based updates
are performed, and then the updated model parameters and optimizer state are flushed back to the third-level storage, while at the same time an FP16 version is pushed to the GPUs. Efficient pipelining and asynchronous I/O techniques are used to overlap computations with transfers between the
GPU memory, host memory and third-level storage using specialized engines such as DeepNVMe~\cite{ren2021zero}. However, this is not
enough to alleviate the limited I/O bandwidth of third-level storage.
Furthermore, compute nodes are equipped with many GPUs that share the
same host memory and third-level storage, resulting in competition for
I/O bandwidth that is not mitigated. Another limitation of state-of-art
approaches is the unnecessary back-and-forth movement between the
host memory and third-level storage due to a suboptimal strategy to
handle gradients and to reuse subgroups already available in the host
memory. \looseness=-1

\paragraph*{\bf Key Insights and Contributions}
In this work, we present \proj, a novel approach that aims to alleviate the aforementioned limitations of state-of-the-art offloading techniques that make use of third-level storage to increase the overall memory capacity and I/O bandwidth. Specifically, we leverage a key observation that external storage of HPC systems, e.g., parallel file systems (PFS) and object stores, are typically underutilized during LLM training
(except for occasional checkpointing~\cite{zhang2022opt}), therefore they can complement
node-local NVMe storage to provide a significant boost of I/O bandwidth. Combined with concurrency control to mitigate contention for I/O bandwidth and a better reuse of cached subgroups from one iteration to another, these ideas enable \proj to achieve a 2.5$\times$ speedup over state-of-art approaches.

We summarize the key contributions as follows:

\begin{enumerate}[topsep=0pt,itemsep=0pt,leftmargin=12pt]
    \item We perform an in-depth characterization of resource utilization during different phases of training when the optimizer state is offloaded to SSDs. In particular, we highlight several fundamental observations for our proposal: the CPU memory, utilized as caching buffers for asynchronous data transfers, experiences thrashing at every iteration; the upscaling and offloading of gradients are redundant; modern PFS capable of delivering several GB/s worth of parallel bandwidth (e.g. using Infiniband) remains unused; CPU cores remain idle due to slow disk writes despite overlapping transfers with CPU-based optimizer updates (\S~\ref{sec:motivation}).

    \item Based on the above characterization, we propose a series of design principles: unified multi-level, multi-path asynchronous offloading using virtual tiers; optimized virtual tier concurrency control for multi-path I/O; cache-friendly ordering of model subgroup processing; and delayed in-place mixed-precision gradient conversion during updates (\S~\ref{sec:design:principles}). These design principles are complemented by an I/O performance model detailed in \S~\ref{sec:design:performance-model}. \looseness=-1

    \item We present \proj, an open-source implementation of the design principles that integrates with existing state-of-the-art LLM training runtimes such as DeepSpeed and Megatron. Our implementation insists on low-level optimizations such as process-exclusive multi-thread-shared locking mechanism in \texttt{libaio}-- an optimized asynchronous POSIX I/O library; and efficient non-sequential ordering of asynchronous prefetch and flush operations to/from the disk (\S~\ref{sec:design:impl}).

    \item We evaluate \proj in a series of experiments in which we pretrain up to 280B parameters model on 32$\times$A100-40GB GPUs. Our approach accelerates both backward and update phases by 13.5$\times$ and 2.3$\times$, respectively, and speeds up the end-to-end training by 2.5$\times$ as compared to the state-of-art DeepSpeed
    (\S~\ref{sec:expt}).
\end{enumerate}

\section{Background and Related Work}
\label{sec:background}

\begin{figure*}
    \centering
    \includegraphics[width=0.9\linewidth]{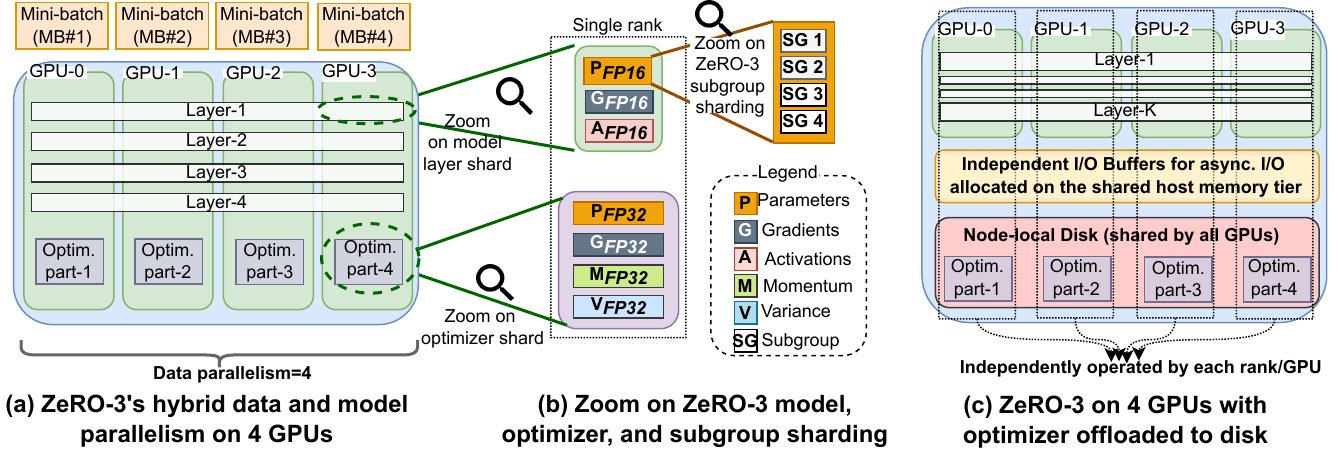}
    \caption{DeepSpeed ZeRO-3 showing (a) hybrid model and data parallelism; wherein (b) the model parameters, activations, gradients, and optimizer states are sharded into subgroups; and (c) the optimizer states are offloaded to node-local NVMe.}
    \label{fig:background}
    \Description{A set of illustrations showing the sharding of model and optimizer states with DeepSpeed's ZeRO-3 approach. Subfigure (a) shows model and optimizer sharded across 4 GPUs when using data parallelism with ZeRO-3 without tensor or pipeline parallelism. Subfigure (b) zooms on the chunks of model and optimizer on each GPU to show parameters, gradients, activations, momentum, variance with their corresponding FP16/FP32 precisions. It then zooms in on each of the above entities being further sharded into subgroups. Subfigure (c) show the offloading of optimizer states on node-local disk.}
\end{figure*}

\paragraph*{\bf Data, Pipeline, and Tensor Parallelism}
Various parallelism techniques for data, pipelines, and tensors, have been widely adopted to accelerate the training of large models, such as Convolutional Neural Networks~(CNNs)~\cite{zhao2024review}, Deep Learning Recommendation Models~(DLRMs)~\cite{zhang2019deep}, Large Language Models~(LLMs)~\cite{zhao2023survey}, and Large Vision Models~(LVMs)~\cite{jiang2024effectiveness}. Data parallelism accelerates training by running multiple model replicas, each running forward and backward passes in parallel with different input mini-batches and synchronizing at the end of the iteration in the update phase to combine the patterns learned from all mini-batches. Pipeline~\cite{huang2019gpipe} and Tensor~\cite{cai2021tensoropt,shoeybi2019megatron} parallelism techniques
split large models across GPUs with limited memory capacities. While pipeline parallelism splits the model vertically by placing a subset of model layers on a given GPU, tensor parallelism
performs horizontal partitioning by splitting the model across all available GPUs. The combination of data, pipeline, and tensor-parallelism, often termed ``3D parallelism'', is used to effectively scale large-scale training across thousands of GPUs for CNNs, LVM, LLMs, DLRMs, etc. However, given the ever-growing large sizes of LLMs and their rapid adoption in various fields, optimizations beyond 3D parallelism are required to run in resource-constrained setups. \looseness=-1

\paragraph*{\bf ZeRO Redundancy Elimination}
State-of-the-art LLM training runtimes, e.g., PyTorch's Fully Sharded Data-Parallel~(FSDP)~\cite{Zhao:FSDP:VLDB2023}, Microsoft DeepSpeed's Zero Redundancy~(ZeRO)~\cite{zero-deepspeed-intro}, Colossal-AI~\cite{li2023colossal}, use redundancy elimination techniques to remove redundant parts of the model and optimizer states across data parallel ranks to minimize GPU memory consumption. To this end, DeepSpeed, a widely used~\cite{workshopBLOOM176BParameterOpenAccess2023,DeepSpeed4Science23} LLM training runtime, proposes three stages for eliminating redundancy across data parallel ranks: ZeRO-1 splits the optimizer states; ZeRO-2 splits both the optimizer states and gradients; and ZeRO-3 splits optimizer states, gradients, and model parameters~\cite{zero-deepspeed-intro}. As illustrated in Figure~\ref{fig:background}(a), training a $P$ parameters model on $N$ GPUs with full redundancy elimination (ZeRO-3) leads to $\sim \lceil \mathcal{O(P)}/N \rceil$ order of memory savings at the expense of 1.5$\times$ higher communication overheads~\cite{zero-deepspeed-intro}. \looseness=-1

The ZeRO-3 technique partitions model states across GPUs, requiring frequent scatter-gather collectives to reconstruct layers on demand, significantly increasing communication costs. Consequently, ZeRO-3, despite its memory efficiency, cannot be seamlessly combined with pipeline parallelism, which relies on efficient inter-stage communication. Instead, ZeRO-3 employs a hybrid model and data parallelism strategy, where training on $N$ GPUs typically results in $N$ ``virtual'' data-parallel replicas. Unlike conventional data parallelism where the model is fully replicated, these replicas remain virtual because model states are dynamically fetched and synchronized across GPUs as required.  \looseness=-1

\paragraph*{\bf Mixed Precision Training}
Mixed precision training, proposed by Baidu and Nvidia research~\cite{micikevicius2018mixedprecisiontraining}, is another widely adopted approach to improve throughput and reduce the memory footprint in LLM training. This is illustrated in Figure~\ref{fig:background}(b) in the left dotted block. Specifically, mixed precision uses two different copies of the model parameters, one in \texttt{FP16} (or \texttt{BF16}), used to run the forward and backward passes, and another master copy in high-precision (\texttt{FP32}), used by the optimizer in the update phase to retain higher stability~\cite{lee2024fp8againquantifyingeffects}. The activations and gradients produced by the forward and backward passes are in half-precision formats, leading to faster communications and faster computations. The low-precision \texttt{FP16} gradients are upscaled to \texttt{FP32} and used by the optimizer to perform the
update phase. Several real-world LLMs, such as BLOOM-176B~\cite{workshopBLOOM176BParameterOpenAccess2023}, OPT-175B~\cite{zhang2022opt}, GPT-3~\cite{Brown:GPT3:NIPS:2020}, and GLM-130B~\cite{zeng2022glm}, are pre-trained using mixed-precision, thereby demonstrating the stability and efficiency of the mixed-precision approach.\looseness=-1

\paragraph*{\bf Sharded Model and Optimizer States Into Subgroups} To reduce the intermediary memory required during computations, DeepSpeed's ZeRO-3 shards the model parameters and optimizer states of each rank/GPU into subgroups, as depicted in Figure~\ref{fig:background}(b). The subgroup sharding technique is unique to the DeepSpeed runtime and is unavailable on other runtimes such as FSDP or Colossal-AI. The size of these subgroups, $M$, is user-defined and specifies the total number of parameters per subgroup.
The subgroups are evenly distributed among the GPUs.
When using mixed precision for training, for each subgroup of $M$ parameters, the forward and backward passes operate on the FP16 parameters and FP16 gradients, while the update step operates on the $M$ corresponding FP32 parameters and FP32 optimizer state. This subgroup-style sharding allows for efficient piecewise computation and communication overlaps while minimizing memory footprint, as explained in ZeRO-Infinity~\cite{ZeroInfinity-SC21}.

\paragraph*{\bf Optimizer State Offloading}
The optimizer state, held in FP32, is much larger than
the model parameters and is only required during the update step. Thus, in memory-constrained scenarios, offloading it
to the host memory (and further other multi-level tiers such as node-local NVMe) is a practical choice. The use of subgroups further facilitates swapping between the host
memory and other multi-level tiers. This is shown in Figure~\ref{fig:background}(c). When optimizer offloading is enabled, updates are typically performed on the CPU because (a) GPU memory is typically fully utilized by FP16 model parameters, and (b) transferring FP32 optimizer states and gradients between GPU and CPU, even subgroup-by-subgroup, incurs high I/O overhead. To efficiently manage these subgroup updates, each GPU worker allocates an independent I/O buffer on shared host memory, with enough capacity to hold a configurable number of subgroups. These I/O buffers serve two primary purposes: (1) caching FP16 gradients during the backward pass, upscaling them to FP32, and asynchronously flushing the FP32 gradients to disk for consumption during the update step, and (2) offloading to disk using asynchronous swap-in/out of optimizer subgroups between the host memory and the disk.

\paragraph*{\bf Accelerating Model and Optimizer Offloading} Several recent efforts focus on mitigating the I/O bottlenecks encountered due to offloading for training, fine-tuning, and inference scenarios. For instance, Fuyou~\cite{liao2024adding} and LoHan~\cite{liao2024lohanlowcosthighperformanceframework} perform pipelined overlapping activation, parameter, and optimizer swapping across GPU-CPU-SSD to accelerate fine-tuning for extremely constrained scenarios, wherein neither model nor optimizer fits on the GPU memory; and is therefore not applicable for the scenario targeted in our work. Smart-Infinity~\cite{jang2024smart} uses computational storage devices (NVMe attached to FPGAs) to mitigate I/O between the NVMe and CPU, which accelerates the update phase but requires specialized NVMe devices. To mitigate the slowdown due to the expensive update phase, asynchronous update techniques such as one-step-delayed updates had been previously proposed in ZeRO-Offload~\cite{ren2021zero}: it overlaps the next iteration on the GPUs (with stale parameters) with the still ongoing CPU-based updates. However, this optimization was later removed from the runtime due
model inconsistencies~\cite{one-step-delayed-deepspeed}. DeepSpeed TwinFlow~\cite{deepspeed-offloadapp} and Deep Optimizer States~\cite{maurya2024deep} are complementary efforts that exclusively focus on optimizing CPU-only offloading and do not optimize disk-based offloading.

\section{Our Contribution: \proj}
\label{sec:design}

\subsection{Gap Analysis}
\label{sec:motivation}

\begin{figure*}[!ht]
    \minipage{0.32\textwidth}
        \centering
        \includegraphics[width=\linewidth]{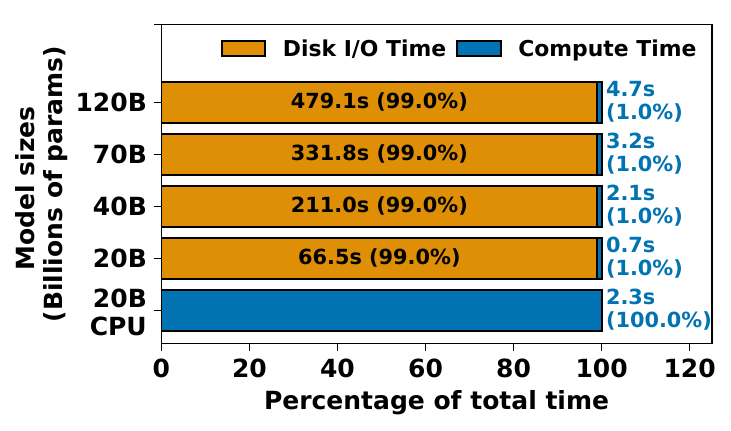}
        \captionsetup{skip=2pt}
        \caption{Fraction of time spent in disk I/O during the update phase}
        \label{fig:motivation-wait-time}
        \Description{A graph showing the breakdown of iteration time, wherein 99\% of time is spent in I/O operations and only 1\% in training computations, when the optimizer is offloaded to the disk.}
    \endminipage
    \hfill
    \minipage{0.32\textwidth}
        \centering
        \includegraphics[width=\linewidth]{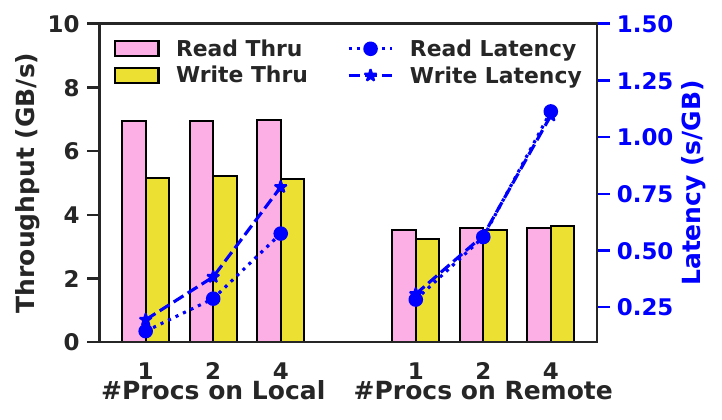}
        \captionsetup{skip=2pt}
        \caption{I/O bandwidth of SSD (local) vs. parallel file system (remote)}
        \label{fig:motivation-disk-thru}
        \Description{A graph showing the I/O bandwidth of the local disk vs parallel file system for increasing number of competing processes within a node. This demonstrates that while the aggregated throughput remains the same, the read/write access latency increases when scaling processes.}
    \endminipage
    \hfill
    \minipage{0.32\textwidth}
        \centering
        \includegraphics[width=\linewidth]{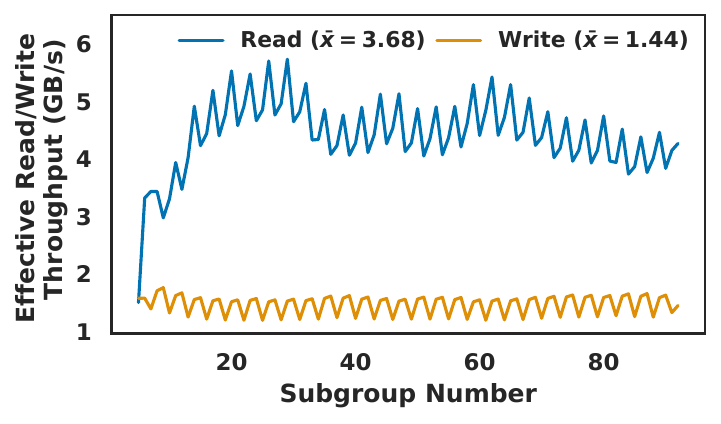}
        \captionsetup{skip=2pt}
        \caption{I/O bandwidth under concurrency per subgroup using a local SSD}
        \label{fig:motivation-per-subgroup-rw-thru}
        \Description{A graph showing the read and write I/O bandwidth fluctions for different subgroups, with mean read and write throughputs at 3.68 GB/s and 1.44 GB/s, respectively.}
    \endminipage
\end{figure*}

The combination of redundancy elimination and multi-tier offloading enables the pre-training and fine-tuning of large LLM models that normally would not fit in the aggregated GPU memory. However, the differences in I/O bandwidth between GPU memory, host memory, and third-level storage are
large, especially when considering node-local SSDs or other forms of
disk storage~\cite{ren2021zero}. As a consequence, I/O overheads may significantly delay
especially the update phase, during which the offloaded subgroups need
to be fetched, updated and flushed back.

\paragraph*{\bf Training Iterations are Dominated by the Update Phase}
As a first step, we study the breakdown of a training iteration in terms
of time spent in the forward pass, backward pass and the update phase, in order
to check whether the update phase occupies a significant fraction of
the overall iteration duration, which would validate the hypothesis
that accelerating the update phase would accelerate the overall training
iteration. To this end, we study DeepSpeed~\cite{rasley2020deepspeed}, configured to use redundancy elimination and multi-tier offloading that includes a local SSD. We run experiments on a single node with 4$\times$H100-80GB GPUs (Table~\ref{tab:testbed}), and record a breakdown of the iterations. We indeed confirmed that the iteration is dominated by the
update phase. For instance, in the case of a 40B model (setup and methodology described in \S~\ref{sec:expt::method}, respectively), the forward and backward passes finish in 0.6s\ (0.02\%) and 28s\ (11\%) of the total 242s\ iteration time, while the update phase finishes in 213s\ (89\%). Thus,
accelerating the update phase is an important step towards a faster
end-to-end runtime.

\paragraph*{\bf Update Phases are Dominated by I/O when using SSD Offloading}
We evaluate four model configurations, i.e., 20B, 40B, 70B, and 120B parameters model (Table~\ref{tab:models}). For the 20B model~\cite{wang2023gemini}, the full FP32 parameters and optimizer state fit in the 512 GB host memory;
therefore, we use it as a baseline. Larger model sizes offload their subgroups to an SSD. As shown in Figure~\ref{fig:motivation-wait-time}, in the case of the 20B model, the update phase completes $\sim$30$\times$ faster than in the case of larger models that offload the model subgroups to an SSD. Specifically, when subgroups are offloaded to the SSD, 99\% of the update phase duration is spent in SSD I/O, despite DeepSpeed’s optimized DeepNVMe engine that
overlaps I/O with CPU computations. Thus, we can conclude that accelerating
the fetches and flushes of offloaded subgroups to/from the host memory
will significantly accelerate the update phase, which according to
the previous experiments, will significantly accelerate the overall
iteration.

\paragraph*{\bf I/O Bandwidth of Different Third-Level Storage}
HPC systems are not only equipped with node-local SSDs but also external storage such as parallel file systems (PFS) and object stores (e.g. DAOS). Although external storage is shared by all compute nodes, and therefore they are subject to I/O competition, at a medium scale, it can provide a significant aggregated I/O bandwidth boost to complement the limited I/O bandwidth of node-local SSDs.
To quantify this opportunity, we measure the raw read and write throughput of both node-local NVMe storage and remote PFS using microbenchmarks. Remote storage can offer higher throughput than node-local NVMe in some cases (e.g., Testbed-2 in Table~\ref{tab:testbed}). Additionally, given that each GPU process independently issues fetch and flush operations for its optimizer state subgroups, understanding I/O performance under contention is crucial. Figure~\ref{fig:motivation-disk-thru} shows that while overall read/write throughputs remain constant as the number of concurrent processes increases, per-process latency (shown using lines and minor y-axis) worsens due to contention within the storage subsystem. This suggests that shared NVMe bandwidth saturation is a limiting factor, making I/O the primary bottleneck at scale.

\paragraph*{\bf I/O Bandwidth under Concurrency during Offloading}
Finally, we analyze the effective read/write throughput perceived by the training runtime when offloading the optimizer states of a 40B model to a node-local NVMe. Figure~\ref{fig:motivation-per-subgroup-rw-thru} reveals oscillations in throughput due to runtime memory constraints, which restrict active subgroups in host memory to three at a time: one prefetched, one actively updated, and one flushed back to disk. The slow flush-back rate results in cases where the next subgroup is prefetched before the previous one is fully written, leading to intermittent spikes in read-throughput. However, the aggregated read and write throughput remains bottlenecked by NVMe’s write bandwidth, reinforcing the I/O limitations observed in prior experiments.

\subsection{Design Principles}
\label{sec:design:principles}

\paragraph*{\bf Unified Multi-level, Multi-path Asynchronous Offloading using Virtual Tiers}
\label{sec:design:principles:multi-tier}
Leveraging local SSDs for offloading is scalable because the SSDs can be leveraged independently using model parallelism to achieve a high aggregated I/O bandwidth. However, model subgroups offloaded to local storage, such
as SSDs, introduce high I/O overheads that dramatically slow down the update phase, as discussed in \S~\ref{sec:motivation}. On the other hand, external storage has significant potential to complement the limited I/O bandwidth of node-local SSDs and is typically under-utilized during pretraining (save for occasional checkpoints)~\cite{XIAN2024103095}. It is this observation that we capitalize on to extend existing multi-level offloading techniques with support for multi-path I/O at each level. Specifically, we unify all alternative storage (local SSDs, parallel file systems~(PFS), object stores) into a virtual, third-level tier that can be used to offload model subgroups from the host memory. Then, based on an I/O performance model (detailed in \S~\ref{sec:design:performance-model}), we assign the subgroups to the alternative storage tiers proportionally to their I/O bandwidth. Using this approach, we can parallelize I/O operations such that slow tiers finish roughly at the same time as fast tiers because they store fewer subgroups. This I/O load balancing allows our approach to avoid any of the tiers becoming a bottleneck, thereby maximizing the acceleration of I/O operations for the overlapping with the computations.
Although we illustrate this principle for the third-level tier, it can be generalized to any level (e.g., second-level tier to combine GPU-CPU or GPU-GPU HBMs, DDR, CXL, etc.).

\paragraph*{\bf Optimized Virtual Tier Concurrency Control for Multi-Path I/O}
\label{sec:design:principles:process-atomic}
To enable high scalability for model parallelism, we allow each worker process (typically attached to a single GPU), to apply I/O load balancing on alternative storage independently of other workers. Using this approach, we avoid the need for expensive global synchronization. However, each compute node features multiple GPUs, which means that multiple processes will compete
for a shared I/O bandwidth to alternative storage tiers. Therefore, we introduce a lightweight concurrency control strategy at the node level
that allows only one worker process on each compute node
to access a given alternative storage at a time. With this restriction of exclusive access
per alternative storage, the full I/O bandwidth is guaranteed to be available to a worker process, which achieves lower latency (Figure~\ref{fig:motivation-disk-thru}) and I/O load balancing. At the same time, the other worker processes
are free to compute updates for prefetched groups using all the available CPU cores, and use other alternative storage(s) in parallel, resulting in a natural interleaving
that achieves global I/O load balancing. Note that a worker is not limited to accessing an
alternative storage using a single I/O thread; it can leverage the preferred I/O parallelism
of the alternative storage (e.g., a PFS is faster when using multiple I/O threads~\cite{VELOC-FGCS24}). Furthermore, if the preferred I/O parallelism cannot be saturated by a single
worker, then the alternative storage can be subdivided further into multiple virtual tiers.

\paragraph*{\bf Cache-Friendly Ordering of Model Subgroup Processing}
\label{sec:design:principles:enable-caching}
An important observation during the update phase of adaptive optimizers (e.g., ADAM~\cite{kingma2014adam}) is that the computations are
embarrassingly parallel: each model subgroup has its own set of corresponding
parameters, optimizer state and gradients that can be applied independently
to obtain the new model parameter and optimizer state. Therefore, \emph{the order in
which the subgroups are independently processed is inconsequential and does not impact accuracy or convergence}. We exploit this observation
to design a cache-friendly update strategy that leverages subgroups present on
fast tiers (host memory) as much as possible. Specifically, in the first iteration, we start processing the subgroups in increasing order of IDs. After the update phase, the subgroups with a high ID will reside on fast (cache) tiers, while the subgroups with a low ID will reside on the slow tiers as they were evicted to slow tiers to make room for the subgroups with higher ID in the host memory. Then, in the second iteration, unlike state-of-the-art approaches,
we reverse the order of processing the subgroups. This results in a
significant acceleration of the update phase because a large number of
subgroups with a high ID are already present on the fast tiers, thereby reducing I/O overheads due to memory thrashing. After the second iteration, the subgroups with low IDs end up on the fast tiers, which means we can take advantage of the increasing order again in the third iteration. Thus, we keep alternating between ascending and descending order of subgroup IDs
which maximizes the benefits of caching.

\paragraph*{\bf Delayed In-place Mixed-Precision Gradient Conversion during
the Update Phase}
\label{sec:design:principles:skip-grads}
As mentioned in \S~\ref{sec:background}, offloading is typically implemented in
combination with mixed-precision training. In this case, state-of-the-art approaches
typically push the FP16 gradients that are computed during the backward pass from
the GPU to the host memory, where they are converted to FP32. From there, they are
flushed to the third-level virtual
tier.
However, during the update phase, each subgroup offloaded to the third-level
virtual tier needs to be brought back to the host memory. Since the subgroup, composed of FP32 optimizer states (parameters, momentum, and variance), becomes even larger because of FP32 gradients, each fetch operation is slower than in the case when the subgroup only holds
the optimizer state.
On the other hand, FP16 to FP32
gradient conversions on a modern CPU have a high throughput (65~GB/s on Testbed-1 in Table~\ref{tab:testbed}) that
is an order of magnitude larger than the fetch throughput. Thus, we propose an alternative
strategy: during the backward pass, we simply store the FP16 gradients on the host memory-- which anyway needs to reserve enough room for the FP16 gradients of all subgroups to enable gradient accumulation.
Then, during the update phase, unlike state-of-the-art approaches, we fetch the subgroup (without FP32 gradients)
and instead convert the FP16 gradients to their FP32 variant on-the-fly (using the same standardized numeric primitives as DeepSpeed~\cite{rasley2020deepspeed, deepspeed-offloadapp}).
Thus, we accelerate both the backward pass (as we eliminate large asynchronous FP32 gradient flushes that can potentially delay the backward pass if they do not
fully overlap with the computations) and the update phase (because the overhead
of in-place FP16 to FP32 gradient conversion is typically negligible
compared with the I/O overhead of fetching FP32 gradients from slow tiers).

\subsection{Performance Model for Subgroup Allocation}
\label{sec:design:performance-model}

To enable load balancing for virtual tiers that can leverage multiple alternative storage options by aggregating multi-path I/O, we propose an approach that assigns model subgroups proportional to the I/O bandwidth of each alternative storage. Specifically, we assume each worker splits its model shard into $M$ equally sized subgroups, which is typically the case to achieve computational load balancing. Furthermore,
we assume a virtual tier composed of $N$ storage tiers $P_i$, where $ 0 \leq i < N$, each with respective I/O bandwidth of $B_i$-- the minimum of read or write throughput. Then, the number of subgroups $T_i$ allocated to
each storage tier $P_i$ can be represented as:

\begin{equation}
T_i = \left\lceil \frac{M \cdot B_i}{\sum_{i = 0}^{N}B_i} \right\rceil, \text{adjusted such that} \sum_{i=0}^{N}T_i = M
\label{eq:perf-model}
\end{equation}

The intuition behind Equation~\ref{eq:perf-model} is to allocate to each alternative storage a number of subgroups roughly equal to the contribution of its I/O bandwidth to the total aggregated I/O bandwidth. This results in parallel fetches and flushes of subgroups from different alternative storage that finish at roughly the same time. Therefore, this will reduce the likelihood of computational stalls due to a straggling
alternative storage while the others remain idle.

Initially, $B_i$ for each alternative storage is measured using microbenchmarks.
Then, after the first iteration, $B_i$ is adjusted based on the average observed I/O bandwidth for subgroup flushes and fetches. This ensures that our approach adapts to any
potential shifts in I/O bandwidth trends that may affect some of the alternative
storage options. For example, a local SSD exclusively owned during a batch job
reservation will not experience I/O bandwidth shifts. However, a parallel file
system may be under I/O pressure from different batch jobs owned by different users,
in which case an updated $B_i$ can modify the value of $T_i$ to repartition the subgroups across different virtual tiers based on their I/O bandwidths.

In addition to faster backward and update phases, the virtual storage tiers in \proj also accelerate the checkpointing process by pre-staging a fraction of optimizer states to persistent storage. This can be leveraged by multi-tier asynchronous checkpointing engines such as DataStates-LLM~\cite{datastates-llm} to flush the remainder of model and optimizer states from the GPU memory, host memory, and the non-persistent storage tiers, such as local-disk, during the immutable forward and backward passes.

\subsection{System Composition of \proj}
\label{sec:design:example}

\begin{figure*}
    \centering
    \includegraphics[width=0.97\linewidth]{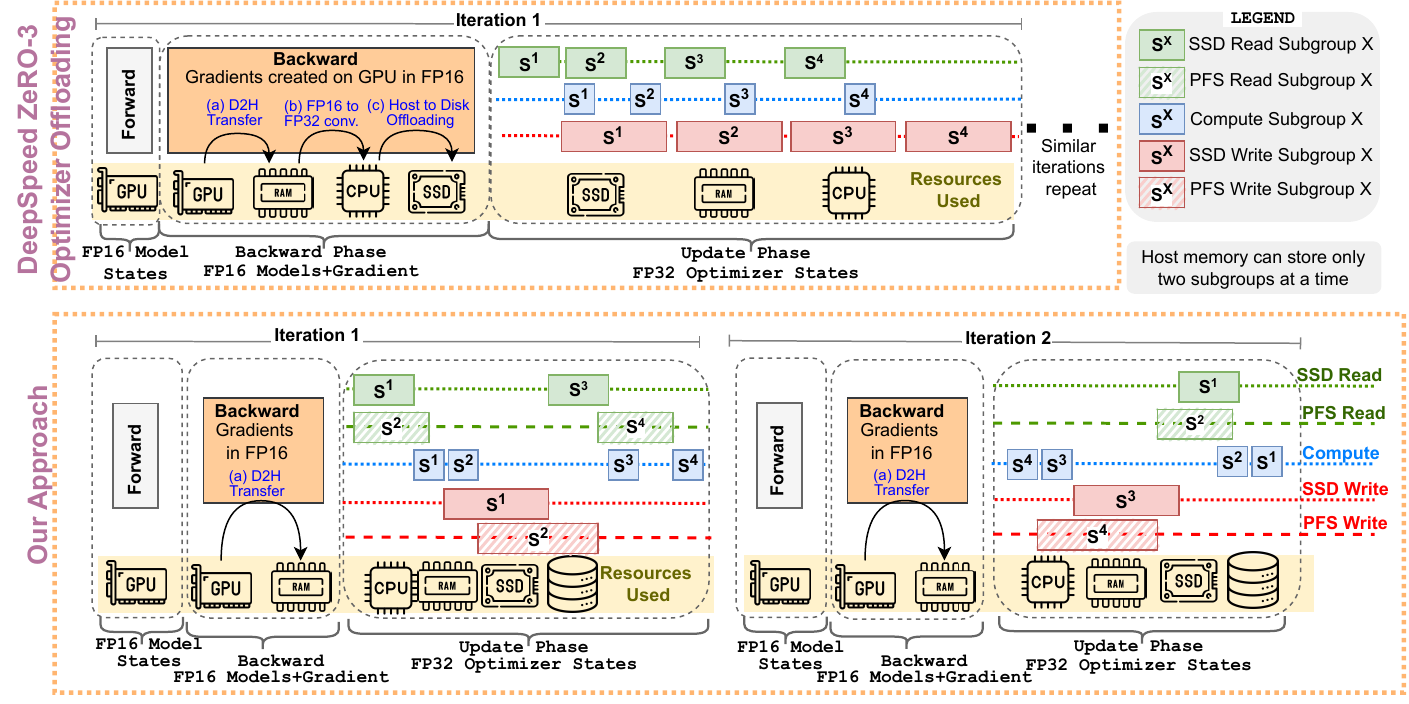}
    \captionsetup{skip=1.5pt}
    \caption{Illustrative example: application of the design principles proposed by \proj vs. state-of-art (DeepSpeed ZeRO-3).}
    \label{fig:multi-path-nvme-arch-update}
    \Description{An illustration showing the working of DeepSpeed ZeRO-3 vs our approach. ZeRO-3 upscales the gradient to FP32 and flushes them to disk, which we eliminate by caching on the CPU memory. Additionally, our cache friendly subgroup reordering, and multi-path offloading to local disk and PFS accelerates I/O, resulting in faster training iterations.}
\end{figure*}

We illustrate how to combine the design principles mentioned above in
Figure~\ref{fig:multi-path-nvme-arch-update}, showing the end-to-end iteration execution of DeepSpeed's ZeRO-3 vs our approach. Specifically, we assume a simplified LLM architecture where each model shard on a GPU is composed of four subgroups denoted by $S^1 \dots S^4$.
The same pattern happens in parallel on the other processes
when using model parallelism. DeepSpeed ZeRO-3 computes
the FP16 gradients for each subgroup during the backward pass, and then
flushes them to the host memory in the background.
On the host, the FP16 gradients are converted to FP32 and flushed to the NVMe. Then, during the update phase, each subgroup (composed of FP32 parameters, momentum, variance, and gradients) is asynchronously fetched from the NVMe using a pipeline: as soon as $S^1$ is available in the host memory, a corresponding parallel multi-core CPU computation is triggered to update the model and optimizer states. Meanwhile, $S^2$ is fetched to the host memory. When the computation is complete,
the subgroup's new parameters are fetched on GPU and the optimizer state is flushed to the SSD after discarding FP32 gradients. The fetches, updates, and flushes overlap throughout the update phase. At the next iteration, the same sequence of patterns repeats.

\begin{algorithm}[t]
    \small
    \caption{Optimizer Multi-Tier Offloading and Updates}
    \label{algo:opt-update}
    \DontPrintSemicolon
    \SetKwInOut{KwIn}{Input}
    \SetKwFunction{RunUpdate}{run\_update}
    \SetKwFunction{AsyncFlush}{async\_h2f\_flush}
    \SetKwFunction{AsyncPrefetch}{async\_f2h\_prefetch}
    \SetKwFunction{AsyncGPUTrf}{async\_h2d\_transfer}
    \SetKwFunction{CPUUpdate}{cpu\_update\_kernel}
    \SetKwFunction{PrefetchWait}{f2h\_prefetch\_wait\_subgrp}
    \SetKwFunction{NextSubgroup}{next\_subgroup}
    \SetKwFunction{AssignTier}{assign\_storage\_tier}
    \SetKwFunction{StorageTier}{storage\_tier}
    \SetKwProg{Fn}{Function}{:}{}

    \KwIn{$iter$: training iteration number; $\langle S \rangle$: Optimizer subgroups; $\langle B \rangle$: I/O bandwidth of local and remote tiers. \texttt{f2h}, \texttt{h2f}, and \texttt{h2d}, denote file-to-host, host-to-file (for NVMe and PFS), and host-to-device (GPU) transfers, respectively.}
    \Fn{\RunUpdate{$iter$, $\langle S \rangle$, $\langle B \rangle$}}{
        $update\_order \gets \{0, 1, \dots |S|-1\}$ or $\{|S|-1, |S|-2, \dots 0\}$\;
        \tcp*{\textcolor{blue}{\scriptsize\ttfamily reversed at each update phase to maximize host memory hits}}
        \For{$i \in update\_order$} {
            $pt \gets \StorageTier(i)$ \;
            $s \gets \PrefetchWait(i)$ \;
            $s.\texttt{grad} \gets \texttt{host\_grad\_accum}_{FP16}[i].to(\mathrm{\texttt{FP32}})$\;
            \CPUUpdate{$s$}\;
            \AsyncGPUTrf{$s.\texttt{params}.to(\mathrm{\texttt{FP16}}$)}\;
            $t \gets \AssignTier(i, B) $ \;
            $\AsyncFlush{s, t}$\;
            $\AsyncPrefetch{\NextSubgroup{i}, pt}$\;
        }
    }
\end{algorithm}

In the case of our approach, we create a virtual third-level tier that combines the local NVMe with a parallel file system (PFS). Initially, the
subgroups are created on the host memory and flushed to either the NVMe
or PFS (according to the performance model discussed in \S~\ref{sec:design:performance-model}). The FP16 parameters are copied
to the GPU. Then, the training iterations can start. Unlike ZeRO-3, the backward pass does not upscale and flush gradients to storage tiers.
Algorithm~\ref{algo:opt-update} provides a high-level pseudo-code description, complementing Figure~\ref{fig:multi-path-nvme-arch-update}. Specifically, during the update phase, $S^1$ and $S^2$ are fetched in parallel from the NVMe and PFS through separate I/O paths. As soon as $S^1$ completes (Line\#5), the corresponding gradients on the host accumulation buffer are upscaled from FP16 to FP32 (Line\#6), after which the update computation starts (Line\#7). Then, an asynchronous host-to-device transfer of the downscaled FP16 model parameters is initiated (Line\#8) after which the new storage tier $t$ for $S^1$ is determined using equation~\ref{eq:perf-model} in Line\#9. Subsequently, the asynchronous flush of $S^1$ to its new tier $t$ and asynchronous prefetch of the next group
on the previous tier $pt$ of $S^1$ is initiated (Lines \#10-11). As seen in Figure~\ref{fig:multi-path-nvme-arch-update}, since the fetch of $S^2$ finishes before the update phase of $S^1$, the update computation of $S^2$ can start immediately after the update computation of $S^1$. Meanwhile, the flush of $S^1$ and prefetch of $S^4$ can progress independently, again on separate I/O paths. While the asynchronous prefetch and flush operations can be enqueued, they might get deferred due to the tier-exclusive concurrency control mechanism (\S~\ref{sec:design:principles}) when competing with other node-local processes. Similarly, $S^3$ and $S^4$ have an opportunity to leverage separate I/O paths for fetches. However, unlike ZeRO-3, note that we do not need to flush $S^3$ and $S^4$ to the virtual tier, because we can switch the order in the next
update phase (Line \#2) to directly use $S^4$ and $S^3$ from
the host memory.

\subsection{Implementation and Integration}
\label{sec:design:impl}

We implement the proposed design principles in \proj~\footnote{\url{https://github.com/DataStates/artifacts/blob/main/MLP-Offload}} as a standalone open-source library integrated with DeepSpeed ZeRO-3. Specifically, the integration extends DeepSpeed’s DeepNVMe offloading engine, which leverages \texttt{libaio}~\cite{libaio}, a kernel-accelerated asynchronous I/O library, to support multi-path parallel read/write operations across processes (each mapped to GPUs) and storage tiers. In addition to benefiting from kernel-accelerated I/O, DeepNVMe’s C++-based implementation avoids inefficiencies introduced by Python’s Global Interpreter Lock (GIL) and PyTorch’s pooled memory management. Building on this foundation, \proj orchestrates efficient host buffer management through explicit pool-based allocations for asynchronous fetch/flush operations, enabling fine-grained concurrency and reducing I/O contention when processes share a storage tier.

\proj can be enabled and configured via two JSON key-value pairs in the DeepSpeed runtime configuration. During ZeRO-3 initialization, we instantiate multiple offloading engine objects per process, corresponding to the number of storage tiers. Each offloading object is assigned a dedicated host buffer to facilitate asynchronous prefetching and lazy flushing. The host buffer size is configurable based on available host memory, the number of processes per node, and the number of storage tiers. Additionally, \proj allows user-specified distribution of subgroups across storage tiers, guided by the performance model (\S~\ref{sec:design:performance-model}). For example, a 2:1 split between \texttt{/local/} and \texttt{/remote/} directories ensures that for every subgroup stored remotely, two are offloaded to the local disk. Beyond managing multiple offloading engines, \proj dynamically predicts subgroup prefetch order and where to lazily flush updated subgroups based on the offloading ratio. While designed for Megatron-DeepSpeed with ZeRO-3, the core principles of \proj make it extensible to other training runtimes, such as TensorNVMe~\cite{TensorNVMe} in Colossal-AI~\cite{li2023colossal} by specifying multiple \texttt{DiskOffloader} objects to create the virtual third-level tier, on each of which the corresponding subgroups dictated by our performance model can be consequently offloaded.

\section{Performance Evaluation}
\label{sec:expt}

\subsection{Methodology}
\label{sec:expt::method}

\paragraph*{\bf Experimental Setup}
We conducted our experiments on Testbed-1 (ANL JLSE)~\cite{argonne_nvidia_h100} and Testbed-2 (ALCF Polaris)~\cite{polaris_alcf},
consisting of 4$\times$H100-80GB and 4$\times$A100-40GB GPUs per node, respectively, outlined in Table~\ref{tab:testbed}. The ratio of host memory to aggregated GPU memory for the Testbed-1 and Testbed-2 platforms are 1.6:1 (similar to AWS \texttt{p4de.24xlarge}) and 3.2:1 respectively (similar to AWS \texttt{p5.48xlarge}), neither adequate enough to hold the 8:1 optimizer to model state memory ratio described in ZeRO-Inifinity~\cite{ZeroInfinity-SC21}, compelling NVMe offloading. Both platforms consist of 2$\times$ RAID-mounted 1.6~TB NVMe M2 SSDs for local storage, the read and write throughputs of which are listed in Table~\ref{tab:testbed}. The Testbed-1 nodes feature 2$\times$ Intel(R) Xeon(R) Platinum 8468, consisting of 96 CPU cores, while the Testbed-2 nodes feature 1$\times$ AMD EPYC 7543P, consisting of 32 CPU cores. We use
Testbed-1 for small-scale experiments and Testbed-2 for scalability experiments. In terms of external storage use as alternative offloading tiers, Testbed-1 features a VAST~\cite{vast2023platform}
parallel file system of 1~PB capacity using 4 DNodes, whose read/write throughputs roughly correspond to speeds of AWS's advanced FSX Lustre FS~\cite{aws_fsx_lustre_pricing}. Testbed-2 is composed of 100~PB storage using HPE ClusterStor E1000 platform through 160 (OSTs),
with read/write throughputs described in Table~\ref{tab:testbed}.

\begin{table}[t]
    \small
    \centering
    \caption{Testbed configurations.}
    \begin{tabular}{|c|c|c|}
         \hline
         Feature $\downarrow$ | Testbed $\rightarrow$ & Testbed-1~\cite{argonne_nvidia_h100}      & Testbed-2~\cite{polaris_alcf} \\
         \hline
         \hline
         GPUs                                   & 4$\times$~H100-80GB    & 4$\times$~A100-40GB \\
         Pinned D$\leftrightarrow$H B/W (GB/s)  & 55                & 25 \\
         Number of CPUs cores                   & 96                & 32 \\
         Per node host memory (GB)              & 512               & 512 \\
         NVMe Read | Write thruput (GB/s)       & 6.9 | 5.3         & 13.5 | 4.8  \\
         Parallel File System~(PFS)             & VAST FS           & Lustre FS \\
         PFS Read | Write throughput (GB/s)        & 3.6 | 3.6         & 6.9 | 13.7   \\
         \hline
    \end{tabular}
    \label{tab:testbed}
\end{table}

\paragraph*{\bf Compared Approaches}
We compare {\bf \proj}, illustrated in Figure~\ref{fig:multi-path-nvme-arch-update}~(bottom) with \textbf{\emph{DeepSpeed ZeRO-3}}, which is a prominent
implementation of the latest state-of-art in LLM training and illustrated in Figure~\ref{fig:multi-path-nvme-arch-update}~(top). Specifically, DeepSpeed ZeRO-3 supports NVMe offloading of the optimizer states, similar to Colossal-AI~\cite{li2023colossal}. We use its asynchronous offloading engine, i.e., DeepNVMe, which overlaps all three operations (fetch, update, flush) to accelerate the update phase at the cost of additional host memory required for asynchronous data movement. Given the fact that the host memory is exhausted while holding a fraction of subgroups, we consider the advanced asynchronous engine as a representative baseline.

\paragraph*{\bf Models and Dataset}
The configurations of models used in our evaluations, which are based on real-world LLM training scenarios, are summarized in Table~\ref{tab:models}.
We do not consider models smaller than 40B because their optimizer states are small enough to fit in the host memory (512~GB).
We use a subset of the OSCAR-en dataset consisting of 79K records, included in the repository of the Bloom model~\cite{workshopBLOOM176BParameterOpenAccess2023}, and use the default LLaMA2 \cite{touvronLlamaOpenFoundation2023} tokenizer for preprocessing the dataset into tokens. Unless otherwise noted, similar to the OPT training configuration~\cite{zhang2022opt}, we set the default sequence length to 2048 and microbatch size to 1 to avoid OOM errors in any configuration.

\begin{table}[t]
    \centering
    \small
    \caption{Models used for evaluations. $N_L$:~Number of layers; $D_H$: Hidden dimensions; $AH$: Attention heads.}
    \label{tab:models}
    \setlength{\tabcolsep}{0.7pt}
    \begin{tabular}{|c||c|c|c|c|c|c|c|}
    \hline
    Model  & 40B\cite{wang2023gemini} & 52B\cite{li2024teleflmtechnicalreport} & 70B\cite{touvron2023llama2openfoundation} & 100B\cite{wang2023gemini} & 120B\cite{taylor2022galacticalargelanguagemodel} & 130B\cite{zeng2023glm130bopenbilingualpretrained} & 280B\cite{rae2022scalinglanguagemodelsmethods} \\
    \hline \hline
    $N_L$  & 128 & 64 & 80 & 124 & 96 & 70 & 72 \\
    $D_H$  & 5120  &  8192 & 8192 & 8192 & 10240 & 12288 & 16384 \\
    $AH$   & 40 & 64 & 64 & 64 & 80 & 96 & 128 \\
    \hline
    \end{tabular}
\end{table}

\begin{figure*}
    \centering
    \minipage{0.32\textwidth}
        \centering
        \includegraphics[width=\linewidth]{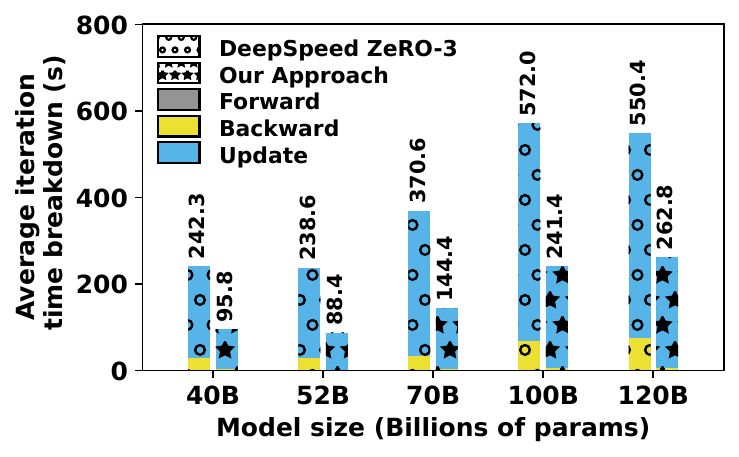}
        \captionsetup{skip=1.7pt}
        \caption{Average iteration time breakdown on scaling model sizes.}
        \label{fig:diff-models-iter-time}
        \Description{A graph showing the average iteration time breakdown on scaling model sizes that demonstrate upto 2.5x speedup with \proj.}
    \endminipage
    \hfill
    \minipage{0.32\textwidth}
        \centering
        \includegraphics[width=\linewidth]{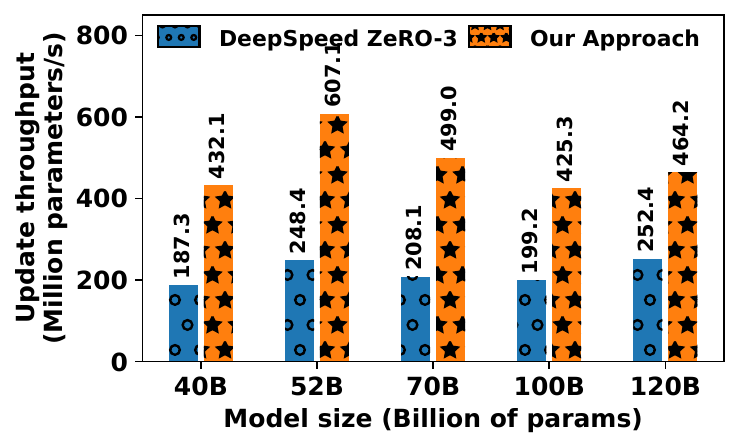}
        \captionsetup{skip=2pt}
        \caption{Average update throughput observed when scaling model sizes.}
        \label{fig:diff-update-throughputs}
        \Description{A graph showing the average update throuhgput observed as millions of parameters updated per second.}
    \endminipage
    \hfill
    \minipage{0.32\textwidth}
        \centering
        \includegraphics[width=\linewidth]{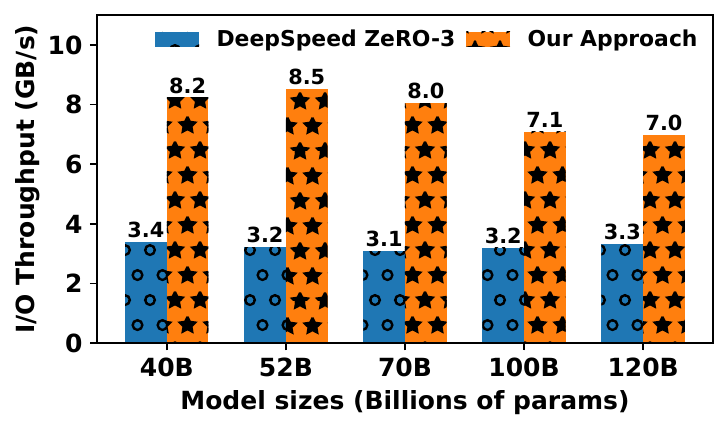}
        \captionsetup{skip=3pt}
        \caption{Effective I/O throughputs for different model sizes.}
        \label{fig:rw-thruputs}
        \Description{A graph showing the effective I/O throughput observed by the training for different model sizes because of caching and multi-path offloading.}
    \endminipage
\end{figure*}

\paragraph*{\bf Runtime Configurations}
As discussed in Section~\ref{sec:background}, DeepSpeed does not support pipeline parallelism in combination with ZeRO-3, which is responsible for the sharding of the model and optimizer states. Therefore, for single-node experiments, we use data-parallelism with ZeRO-3, which shards the model parameters, gradients, and optimizer states to fit in the GPU memory. For weak scalability experiments, we use a combination of tensor-parallelism (intra-node) and data-parallelism (inter-node) approaches to maximize performance and memory savings~\cite{shoeybi2019megatron, ZeroInfinity-SC21}.

In each experiment, all the GPUs on the selected node(s) are utilized, and each GPU is associated uniquely with a single process. For all models, local NVMe offloading is enabled, based on the asynchronous approach for overlapping fetch, flush, and update operations of subgroups. Our approach has additional access to the PFS. To facilitate prefetching and lazy-flushing of subgroups for asynchronous offloading, a configurable number of pinned host buffers are pre-allocated such that the host memory is utilized to the maximum extent ($>$90\% memory utilization) for all compared approaches. The size of this host buffer for asynchronous offloading varies between different models because each model reserves a different amount of runtime-level buffers, such as for gradient accumulation, all-reduce, etc. We refer the reader to ZeRO-Offload~\cite{ren2021zero}, ZeRO-Infinity~\cite{ZeroInfinity-SC21}, and DeepSpeed memory estimator~\cite{deepspeed_memory} for the breakdown of GPU/host memory consumed by the runtime.

Throughout our evaluations, we ensure that the aggregated GPU memory is sufficient to store the following: (1) FP16 model parameters; (2) activation checkpoints generated by the forward pass; and (3) FP16 gradients of one at least subgroup generated during the backward pass which is flushed asynchronously to the gradient accumulation buffer residing on the host memory. We also ensure that the host memory is large enough to hold runtime-level buffers (e.g., gradient accumulation, all-reduce buckets, etc.~\cite{deepspeed_memory}), and a minimum of three subgroups to facilitate asynchronous updates: the previous subgroup being lazily flushed to disk, the current subgroup being updated, and the next subgroup being prefetched from local NVMe.

Although the subgroup sizes do not impact the iteration duration, convergence, or accuracy as mentioned in~\cite{jang2024smart,maurya2024deep}, smaller subgroups achieve better I/O and compute overlap of offloaded subgroups. Therefore, for all approaches, we use a subgroup size of 100 million trainable parameters as opposed to DeepSpeed's default size of 1 billion parameters per subgroup, which allows better
load balancing for our approach. Similar to Turing-NLG 17.2B, GPT-3 175B, BLOOM-176B~\cite{workshopBLOOM176BParameterOpenAccess2023}, we used activation checkpointing to reduce the GPU memory utilization at the expense of 33\% additional recomputations during the backward pass~\cite{ZeroInfinity-SC21}.
This is a popular choice for training with scarce GPU memory. Nevertheless, our approach is complementary to activation checkpointing and would produce similar results without activation checkpointing as well.

\paragraph*{\bf Key Performance Metrics}
We use the following metrics for evaluating the aforementioned approaches:
(1) average time to run a single training iteration (broken down by the duration of the forward pass, backward pass, and update phase); (2) update throughput (expressed as millions of parameters updated per second); (3) effective read/write throughput observed while fetching and flushing disk-offloaded subgroups; and (4) distribution of the optimizer states across different tiers. These metrics are important to understand the end-to-end performance and scalability of LLM training using our approach vs. state of art, as well as to highlight important intermediate steps that influence the end-to-end results.

\begin{figure*}
    \centering
    \minipage{0.32\textwidth}
        \centering
        \includegraphics[width=\linewidth]{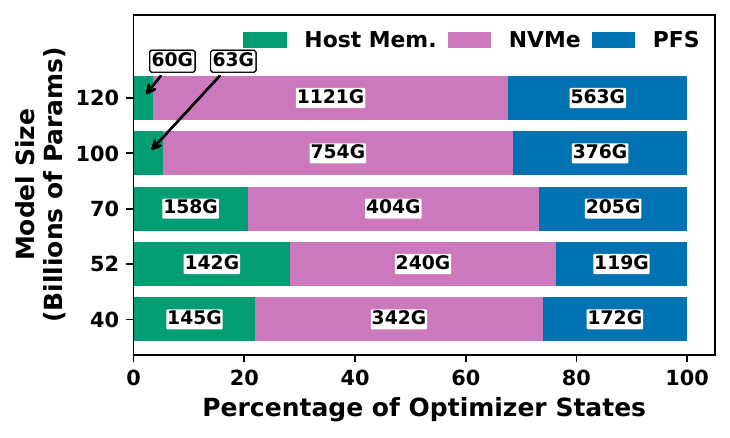}
        \captionsetup{skip=2pt}
        \caption{Distribution of optimizer states across different tiers.}
        \label{fig:optimizer-distribution}
        \Description{A graph showing the distribution of optimizer states cached/offloaded across the host memory, local disk and PFS.}
    \endminipage
    \hfill
    \minipage{0.32\textwidth}
        \centering
        \includegraphics[width=0.95\linewidth]{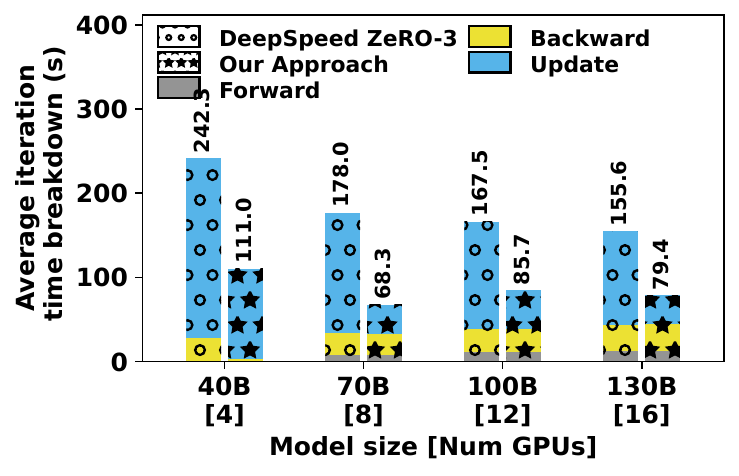}
        \captionsetup{skip=0.4pt}
        \caption{Weak scaling: Iter. time for increasing model sizes with \# GPUs.}
        \label{fig:scalability}
        \Description{A graph showing the iteration time when studying weak scalability by increasing model sizes nearly proportionally to the number of GPUs}
    \endminipage
    \hfill
    \minipage{0.32\textwidth}
        \centering
        \includegraphics[width=0.97\linewidth]{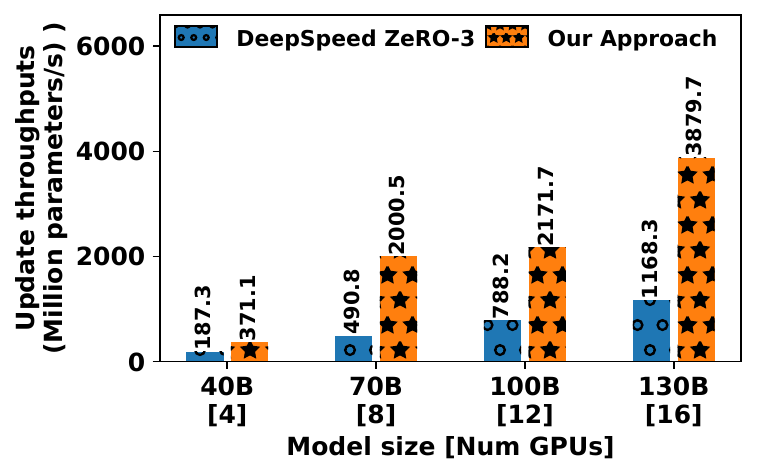}
        \captionsetup{skip=0.4pt}
        \caption{Weak scaling: Update thru. for increasing model sizes with \# GPUs.}
        \label{fig:update-throughputs-scaling}
        \Description{A graph showing the update throughputs observed in the aforementioned weak scaling experiments.}
    \endminipage
\end{figure*}

\subsection{Results: Model Size Scalability}
We first measure the iteration time, broken down by forward, backward, and update phase durations for increasing model sizes as listed in Table~\ref{tab:models} on a single 4$\times$H100 node of Testbed-1 (Table~\ref{tab:testbed}). Each experiment runs for 10 iterations, of which the first 2 are warmups, and the average from 8 iterations is reported. We vary the model size between 40B and 120B such that
all FP16 parameters and gradients for a single subgroup fit within the aggregated 320~GB of GPU memory. With increasing model size, the intensity of offloading increases as well, thus highlighting the effectiveness of the compared approaches as the offloading pressure increases (at 120B parameters, the optimizer state reaches 1.8~TB).

As observed in Figure~\ref{fig:diff-models-iter-time}, the iteration duration follows an increasing trend for increasing model sizes (with slight exceptions for 52B vs. 40B and 120B vs. 100B, as they have fewer transformer layers but more hidden dimensions). As expected, due to offloading, the update phase
is the longest, while the forward pass is almost negligible in comparison.
For DeepSpeed ZeRO-3, the backward pass begins to be noticeable, while our
approach reduces it to a negligible level. We also observe that our approach accelerates the update phase by up to 2.4$\times$, leading to iterations that are overall 2.7$\times$ faster compared with DeepSpeed ZeRO-3.

To further explain these results, we depict in Figure~\ref{fig:diff-update-throughputs} the update throughput, which for reference is $\sim$40000M params/s on the GPUs and $\sim$~8000M params/s on the CPUs when the model parameters and optimizer states are fully available in the GPU and host memory, respectively. With offloading the update throughput drops by an order of magnitude, even on the CPUs, which confirms that the bottleneck is not
on the compute side, but rather due to the slow I/O to the NVMe and/or PFS.
Furthermore, the update throughput stays relatively stable for all model sizes (the update throughput per subgroup remains unchanged, only the number of subgroups changes for different models). Overall, \proj achieves an update throughput 1.8$\times$--2.4$\times$ higher than DeepSpeed ZeRO-3.

\begin{figure*}
    \minipage{0.32\textwidth}
        \centering
        \includegraphics[width=\linewidth]{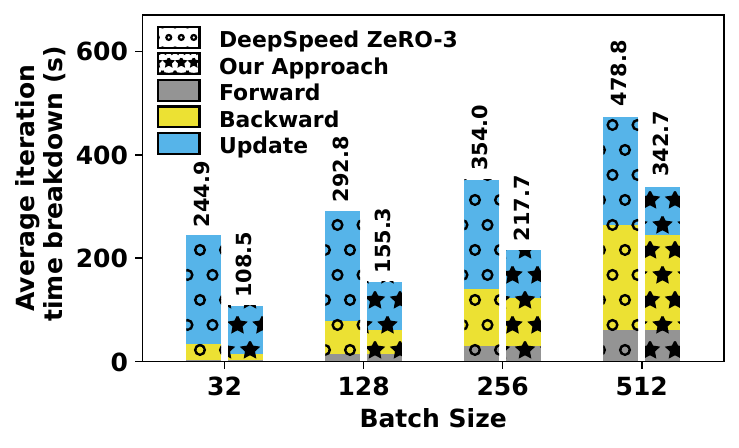}
        \captionsetup{skip=1.8pt}
        \caption{Average iteration time of different batch sizes for the 40B model.}
        \label{fig:diff-grad-accum-iter-time}
        \Description{A graph showing the iteration time of different batch sizes when running 40B model, wherein \proj outperforms ZeRO-3 at increasing batch size scale.}
    \endminipage
    \hfill
    \minipage{0.32\textwidth}
        \centering
        \includegraphics[width=\linewidth]{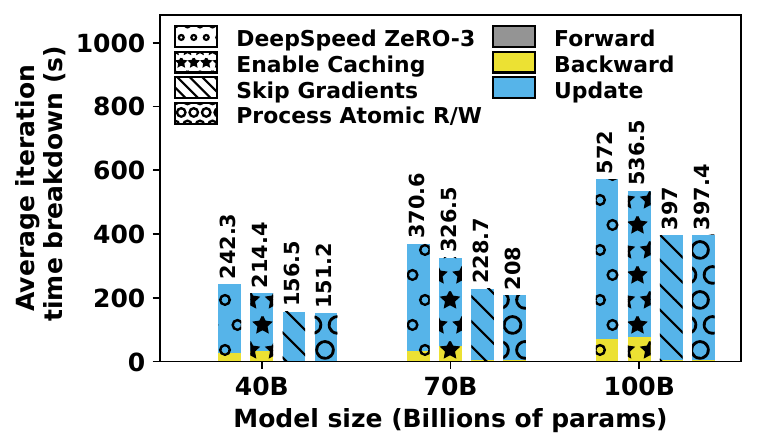}
        \captionsetup{skip=2pt}
        \caption{Performance ablation on node-local NVMe.}
        \label{fig:ablation-local-only}
        \Description{A graph showing ablation experiments to understand the contribution of each design principle when offloading only to node-local NVMe.}
    \endminipage
    \hfill
    \minipage{0.32\textwidth}
        \centering
        \includegraphics[width=\linewidth]{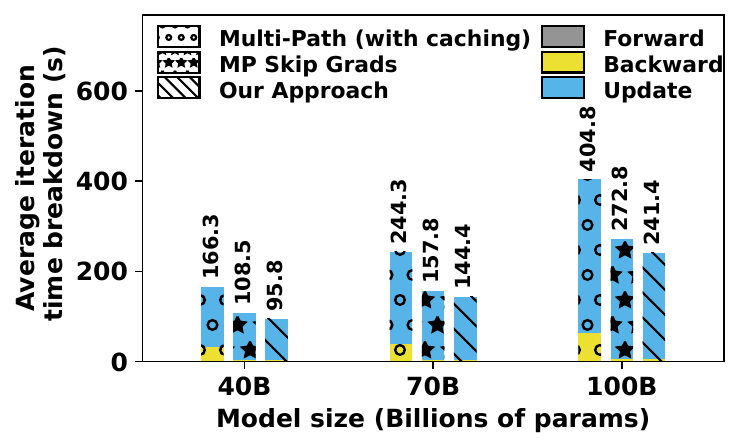}
        \captionsetup{skip=2pt}
        \caption{Performance ablation on node-local NVMe and PFS.}
        \label{fig:ablation-local-vast}
        \Description{A graph showing the performance ablation when offloading to both node-local disk and PFS tiers.}
    \endminipage
\end{figure*}

\subsection{Results: I/O and Storage Tier Load}
As mentioned previously, the update throughput is subject to I/O
bottlenecks. Therefore, we discuss next the I/O throughput sustained
by \proj vs. DeepSpeed ZeRO-3. As opposed to the microbenchmarks discussed
in Figure~\ref{fig:motivation-disk-thru}, running an end-to-end training
overlaps I/O (asynchronous prefetch and flush) with computations, which introduces additional overheads. Specifically, the I/O throughput is computed  as $2\times subgroup\_size\_bytes/(read\_time + write\_time)$, averaged over all subgroups. The size is doubled because every subgroup needs to be both
read and written. Figure~\ref{fig:rw-thruputs} depicts the aggregated I/O throughput of all subgroups for an increasing model size. Interestingly, the DeepSpeed ZeRO-3 approach demonstrates I/O throughput of $\sim$~3.2~GB/s, which is much lower than the peak write speed (5.3~GB/s) of the NVMe on
Testbed-1~(Table~\ref{tab:testbed}). This is because parallel multi-threaded reads and writes from all processes create contention on the CPU-NVMe interconnect (which is PCIe in this case) and on the NVMe storage subsystem. Conversely, \proj alleviates the pressure on the local NVMe by using a PFS, thus benefiting from both multi-tier offloading and better utilization
of individual tiers, ultimately being 2.6$\times$ faster. The effective I/O throughput with \proj decreases slightly with increasing model size because smaller fractions of the optimizer states can be cached on the host memory, thereby reducing the effectiveness of caching. However, \proj still achieves $\sim$2$\times$ I/O speedup for larger models compared to the DeepSpeed ZeRO-3 approach.

To explain the impact of multi-tier offloading, we highlight the distribution of the optimizer states across the different storage tiers. A major chunk of the host memory is consumed by the DeepSpeed runtime for setting up the ZeRO-3 specific data structures that consume 250-350 GB of host memory, proportional to the model size, as reported by DeepSpeed's memory estimator~\cite{deepspeed_memory}. The remaining host memory is used for caching subgroups and asynchronous I/O operations. Figure~\ref{fig:optimizer-distribution} depicts the percentage of optimizer states distributed across the host memory, local NVMe, and the PFS at every iteration. Unlike \proj, DeepSpeed ZeRO-3 experiences cache thrashing on the host buffers due to sequential-ordered subgroup updates and does not utilize the PFS, forcing all optimizer states to be read and written back to local NVMe in each iteration. The fraction of the optimizer states distributed across the local NVMe and the PFS confirm the effectiveness of our performance model (\S~\ref{sec:design:performance-model}), showing a 2:1 NVMe to PFS offloading that is consistent with the read and write throughputs in Table~\ref{tab:testbed}.

\subsection{Results: Weak Scalability}
Next, we study the weak scalability of \proj by varying the model sizes proportionally to an increasing number of nodes. As discussed in \S~\ref{sec:background}, DeepSpeed does not implement pipeline parallelism in combination with the sharding of model parameters and optimizer state. Therefore, we use tensor parallelism across the four co-located GPUs on the same compute node, and data parallelism between the compute nodes.

ZeRO-3's parameter sharding across data-parallel ranks requires frequent scatter collectives leading to higher communication costs during forward and backward passes at the expense of memory savings. Therefore, this scalability study is important to understand if higher communication costs offset the gains achieved by \proj in backward and update phases at scale.

For this experiment, we scale up to 8 nodes (32$\times$ A100-40GB GPUs) on Testbed-2 (Table~\ref{tab:testbed}). The largest model that fits within the aggregated GPU memory in FP16 format is selected from Table~\ref{tab:models}, ensuring a proportional increase in model size with node count: 40B (1 node), 70B (2 nodes), 100B (3 nodes), 130B (4 nodes), and 280B (8 nodes). Figure~\ref{fig:scalability} shows that iteration time decreases for an
increasing number of GPUs. In this case, the communication overheads in forward and backward passes are not as significant compared to the I/O bottlenecks of offloaded optimizer state updates due to the fast interconnect between the compute nodes (e.g., Slingshot, Infiniband) that are typically available on HPC infrastructures. Furthermore, an increasing node count enables independent I/O to the local NVMe, which accelerates the subgroup updates. Consequently, \proj achieves up to 2$\times$ faster iterations than DeepSpeed ZeRO-3, even at scale. To explain this trend, we analyze the update throughputs in Figure~\ref{fig:update-throughputs-scaling}. As expected, update throughput scales with increasing resources (I/O bandwidth and CPUs); and when correlated with the iteration duration this confirms that I/O is still the bottleneck.

In addition to demonstrating \proj's scalability gains over ZeRO-3 with increasing model sizes, the weak scalability experiments also highlight the cost-effectiveness of NVMe-offloading with \proj as compared to the GPU-only training scenario. For instance, training the 70B model without offloading requires the aggregated memory of $\sim$80 A100-40GB GPUs~\cite{datastates-llm} and runs a single iteration in 24s. Conversely, as shown in Figure~\ref{fig:scalability}, with NVMe-based optimizer offloading, ZeRO-3 can run using 10$\times$ fewer GPUs, but takes 168s per iteration, i.e., 7$\times$ slower. In contrast, \proj is only 4.8$\times$ slower, thereby achieving a 5$\times$ slowdown while using 10$\times$ fewer GPUs-- yielding a 2$\times$ improvement in cost-effectiveness compared to GPU-only training.

\subsection{Results: Gradient Accumulation Scalability}
\label{sec:exp:grad-accum}

We next study the impact of using gradient accumulation, which is a popular technique~\cite{workshopBLOOM176BParameterOpenAccess2023} to reduce the
number of update phases (and thus the impact of offloading) by running
multiple forward and backward passes before each update phase. While the number of iterations does not decrease, it is equivalent to running
the training with larger mini-batches (an alternative that is not possible when the GPU memory is scarce). The goal is to show that despite the reduced
frequency of the update phase, our approach still delivers a significant end-to-end speedup compared with the state-of-the-art.

We experiment with the 40B model running on Testbed-1 (4$\times$H100 GPUs),
which can accommodate a mini-batch size of 8 samples, beyond which we encounter out-of-memory on GPUs. Consequently, when running across 4 GPUs in a data-parallel fashion, when we run an update phase for and increasing number of forward and backward passes (1--16), the equivalent batch size increases
in the range of 32--512. As observed in Figure~\ref{fig:diff-grad-accum-iter-time}, even when gradient accumulation is used to amortize the cost of expensive update phases, \proj still outperforms DeepSpeed ZeRO-3 by at least 40\%.

\subsection{Results: Ablation Study}
\label{sec:exp:ablation}

In the last set of experiments, we perform ablation studies to understand the impact of each optimization proposed as part of our design principles (\S~\ref{sec:design}). We consider three models, i.e., 40B, 70B, and 100B, which represent small-scale, medium-scale, and large-scale models, respectively, on Testbed-1. Figure~\ref{fig:ablation-local-only} depicts the accumulated impact (i.e. progressive activation) of each optimization when the optimizer state is only offloaded to the node-local NVMe. The approaches labeled as \textit{Enable Caching}, \textit{Skip Gradients}, and \textit{Process Atomic R/W} correspond to the design principles discussed in ~\S~\ref{sec:design:principles}: cache-friendly subgroup reordering, delayed in-place mixed-precision gradient conversion, and optimized virtual tier concurrency control, respectively. As can be observed, progressive activation of each optimization further reduces the iteration duration, which means
each optimization individually contributes to the speedup, resulting in up to 1.6$\times$ speedup vs. DeepSpeed even without a PFS. Figure~\ref{fig:ablation-local-vast} depicts the same accumulation of optimizations but with the PFS
active (multi-path). In this case, activating all three optimizations is equivalent to our approach. Compared with Figure~\ref{fig:ablation-local-only},
the multi-path parallel I/O further speeds up the iteration by 1.6$\times$, resulting in 2.5$\times$ faster iterations compared with DeepSpeed ZeRO-3.

\section{Conclusions}
\label{sec:conclusion}

In this paper, we present a novel technique, \proj, and its implementation as a library that can be integrated with state-of-the-art LLM runtimes that enable scalable training and fine-tuning.
Specifically, we target the offloading of the optimizer state to a multi-level, multi-path
memory and storage hierarchy to accelerate the training of large LLMs under
GPU memory constraints. In this context, state-of-art approaches suffer from significant
I/O bottlenecks with optimizer state offloading to storage tiers
due to the large size of the full-precision optimizer states (8$\times$ larger than
FP16 parameters), which spill beyond the capacity of the host memory (that is typically only
2$\times$ larger than the aggregated GPU memory) and therefore need to be offloaded to tertiary storage tiers (e.g., node-local NVMe devices), whose I/O bandwidth is orders of magnitude lower.
To reduce these I/O bottlenecks, \proj proposes several design principles, such as multi-level multi-path asynchronous offloading, concurrency control for multi-path I/O, cache-friendly subgroup update reordering, and dynamic in-place mixed-precision gradient conversion. The design principles are implemented as a modular extension to DeepSpeed's offloading engine. Extensive evaluations on 40B--280B parameter models demonstrate 2.5$\times$ faster training iterations as compared to DeepSpeed ZeRO-3 for different configurations at scale. Encouraged by these results, we next plan to explore parallel I/O paths for next-generation Compute-Express Link~(CXL) memory pools and the integration of \proj with other offloading runtimes, frameworks, and accelerators. Of particular interest is a deeper study on the behavior of globally shared alternative storage tiers under I/O competition, which is the case of parallel file systems and object stores. In this case, we plan to explore how to
mitigate predictable fluctuations in I/O bandwidth.

\begin{acks}
This work is supported in part by the U.S. Department of Energy (DOE), Office of Advanced Scientific Computing Research (ASCR) under contract DEAC02--06CH11357/0F--60169 and the National Science Foundation (NSF) under award no.\  2411386/2411387, 2106635. Results presented in this paper are obtained using Argonne's HPC systems-- ALCF Polaris, Joint Laboratory for System Evaluation (JLSE), and NSF's CloudLab and Chameleon testbeds.
\end{acks}

\balance
\bibliographystyle{ACM-Reference-Format}
\bibliography{references}

\end{document}